\documentclass[default,iicol]{sn-jnl}

\usepackage{balance} 

\theoremstyle{thmstyleone}%
\theoremstyle{thmstyletwo}%

\theoremstyle{thmstylethree}%

\begin{document}

\title[Article Title]{Adaptive List Decoder with Flip Operations for Polar Codes}


\author[1]{\fnm{Yansong} \sur{Lv}}\email{lys\_communication@cuc.edu.cn}
\author*[2]{\fnm{Hang} \sur{Yin}}\email{yinhang@cuc.edu.cn}
\author[2]{\fnm{Zhanxin} \sur{Yang}}\email{yangzx@cuc.edu.cn}
\author[3]{\fnm{YuHuan} \sur{Wang}}\email{wangyuhuan@bit.edu.cn}
\author[1]{\fnm{Jingxin} \sur{Dai}}\email{daijingxin@cuc.edu.cn}
\author[1]{\fnm{Jing} \sur{Huan}}\email{huanjing@cuc.edu.cn}

\affil[1]{\orgdiv{State Key Laboratory of Media Convergence and Communication}, \orgname{Communication University of China}, \orgaddress{ \postcode{100024}, \state{Beijing}, \country{China}}}
\affil*[2]{\orgdiv{Engineering Research Center of the Digital Audio and Video Ministry of Education}, \orgname{Communication University of China}, \orgaddress{ \postcode{100024}, \state{Beijing}, \country{China}}}
\affil[3]{\orgname{Beijing Institute of Technology}, \orgaddress{ \postcode{100081}, \state{Beijing}, \country{China}}}


\abstract{Successive cancellation list decoders with flip operations (SCL-Flip) can utilize re-decoding attempts to significantly improve the error-correction performance of polar codes. However, these re-decoding attempts result in extra computation complexity, which thus leads to increased energy consumption and decoding latency to the communication system adopting SCL-Flip decoders. To significantly reduce the computation complexity of current SCL-Flip decoders, we design a new adaptive SCL-Flip (AD-SCLF) decoder, which can be easily implemented based on existing SCL-Flip techniques. Simulation results showed that the AD-SCLF can reduce up to 80.85\% of the computational complexity of a current SCL-Flip decoder at a matched $FER=10^{-3}$. The result implies our decoder can significantly reduce the energy consumption caused by redundant re-decoding attempts from the SCL-Flip decoder.}




\maketitle

\section{Introduction}\label{sec1}

The polar code was proved to reach the channel capacity by utilizing the successive cancellation (SC) decoder with infinite code lengths \cite{[1]}. However, at the code lengths of interest, the SC decoder falls short in providing a reasonable error-correction performance.  Thus, SC-list (SCL) decoder, proposed in \cite{[2]} and \cite{[3]}, improves its error-correction performance by adopting a list of multiple candidate paths. Especially, the improved SCL decoder concatenated with a cyclic redundancy check (CRC), called CRC-aided SCL (CA-SCL) decoder \cite{[2],[4]} can obtain a comparable error-correction performance of current LDPC code at short to moderate code lengths.   But multiple candidate paths bring a larger space and computation complexity than the single path in SC decoding.

To reduce the computation complexity of CA-SCL decoder, many improved decoders were designed for polar codes. Reference \cite{[5]} proposed an adaptive CA-SCL (AD-SCL) decoder with an adaptive list size, which significantly reduces the computation complexity of CA-SCL at medium to high signal to noise ratio (SNR). However, the extra re-decoding attempts introduced by AD-SCL decoder is unfriendly to decoding latency. Therefore, some segmented CA-SCL decoders \cite{[6],[7],[8],[9],[10]} without re-decoding attempts were successively proposed, which reduce the computation complexity of CA-SCL by terminating wrong decoding process early. Similarly, fast list decoders \cite{[11],[12]} also can reduce the complexity by simplifying the recursive computation in decoding processing. These methods all can significantly reduce the computation complexity of the CA-SCL decoder with a similar error-correction performance. 

To further improve the error-correction performance of CA-SCL decoder, \cite{[15]} designed an SCL bit-flip decoder, which firstly introduces flip operations to CA-SCL decoder. \cite{[16]} further improved the efficiency of flip operations and obtained better error-correction performance than that in \cite{[15]}. Moreover, \cite{[17]} improved the efficiency of flip operations in \cite{[16]} by optimizing the order of performing flip operations and further reduced the computation complexity by introducing an adaptive scheme. Meanwhile, \cite{[18]} proposed a new threshold considering the effect of eliminated candidate paths and obtained a similar error-correction performance as \cite{[16]} with a reduced computation complexity. For convenience, these list decoders in \cite{[15],[16],[17],[18]} will be called SCL-Flip decoders in the following sections. 

Although these SCL-Flip decoders improve the error-correction performance of CA-SCL, the introduced flip operations bring a significantly increased number of re-decoding attempts, especially at low to medium SNR. Simultaneously, these re-decoding attempts will increase the energy consumption of the communication system adopting SCL-Flip decoders and require an excessive decoding latency. Therefore, efficiently reducing the computation complexity caused by redundant re-decoding attempts without error-correction performances loss becomes a key to expanding the applied scope of SCL-Flip decoders.

In this work, we designed an adaptive SCL-Flip (AD-SCLF) decoder to significantly reduce the computation complexity of current SCL-Flip decoders at a practical frame error rate (FER). Firstly, we introduce the detail of existing SCL-Flip decoders and obtain indicators of affecting their computation complexity. Then, based on optimizing for some indicators, we proposed an AD-SCLF decoder that can be easily implemented based on existing SCL-Flip decoders. Simulation results showed that our AD-SCLF decoder reduces 80.85\% of the computation complexity of the SCL-Flip decoder in \cite{[18]} at a matched practical {$FER=10^{-3}$}.

The structure of the work is organized as follows. In Section 2, the polar encoding and the detail of decoders related with our work are introduced. In Section 3, the details of the proposed decoder are described. In Section 4, the simulation results are illustrated and discussed. Finally, some conclusions are highlighted.

\section{Preliminaries}\label{sec2}
\subsection{Polar Encoding}
A polar code $PC(N,K)$ of code length $N$ and rate $R=K/N$ are encoded through the following matrix multiplication:
\begin{equation}
	x_{1}^{N}=u_{1}^{N}G_{N},
\end{equation}
where $x_1^N=(x_1,x_2,…,x_N)$ refers to the encoded vector,
$u_1^N=(u_1,u_2,…,u_N)$ represents the encoding vector and $u_i$ is the value of $i^{th}$ bit-channel. $G_N$ is the generator matrix of polar codes.  $G_N$ satisfies
\begin{equation}
	G_{N}=B_{N}F^{\otimes n},
\end{equation}
where $B_N=R_N (I_2 \otimes B_{N/2})$ is a bit-reversal permutation matrix, $R_N$ is an odd-even separation matrix, $\otimes$ refers to Kronecker product and $F=\begin{bmatrix} 1 & 0 \\ 	1 & 1 \\ \end{bmatrix}$. $u_1^N$ can be divided into two subsets: the one containing the frozen bits denoted by $\mathcal{A}^c$ and the other including the non-frozen bits denoted by $\mathcal{A}$.

\subsection{CA-SCL Decoder for Polar Codes}
CA-SCL decoder outputs the final path by utilizing a CRC and the path metric (PM) value proposed in \cite{[19]}.  If the CRC fails at the end of decoding, CA-SCL outputs the estimated codeword with the smallest PM among $L$ candidate paths. $L$ is the list size. If not, CA-SCL outputs the estimated codeword with a successful CRC and the smallest PM. The recursive formula of PM adopted by this work is defined as
\begin{equation}
		{PM}_{l}^{(i)}= \left\{\begin{matrix} 
	{PM}_{l}^{(i-1)},  & \text{if } \hat{u}_{i}[l]=\delta (L_{N}^{(i)}[l])      \\
               {PM}_{l}^{(i-1)}+ \lvert L_{N}^{(i)}[l]\rvert , & \text{otherwise}   \end{matrix}\right.
\end{equation}
where
\begin{equation}
	\delta(x)=\frac{1}{2}(1-sign(x)) \\
\end{equation}
and
\begin{equation}
	sign(x)=\left\{\begin{matrix}1,  & \text{if}\quad x>0 \\
              0,  & \text{if}\quad x=0 \\
              -1, & \text{if}\quad x<0 \\
	\end{matrix}\right.
\end{equation}
$l$ represents the index of the $l^{th}$ candidate path and $ 1 \leq l\leq L$. The initial value of PM is ${PM}_l^{(0)}=0$. $\hat{u}_i$ represents the estimated value of $u_i$. The logarithmic likelihood ratio (LLR) of the $i^{th}$ bit of the $l^{th}$ candidate path is represented by $L_{N}^{(i)}[l]$. The recursions of LLR are given as
\begin{equation}
	\begin{split}
		&L_{N}^{(2i)}(y_{1}^{N},\hat{u}_1^{2i-1}) = \\
		&g(L_{\frac{N}{2}}^{(i)}(y_{1}^{\frac{N}{2}},\hat{u}_{1,o}^{2i-2}\oplus \hat{u}_{1,e}^{2i-2}),L_{\frac{N}{2}}^{(i)}(y_{\frac{N}{2}+1}^{N}, \hat{u}_{1,e}^{2i-2}),\hat{u}_{2i-1})
	\end{split}
\end{equation}
\begin{equation}
	\begin{split}
		L_{N}^{(2i-1)}&(y_{1}^{N},\hat{u}_1^{2i-2})= \\
		&f(L_{\frac{N}{2}}^{(i)}(y_{1}^{\frac{N}{2}},\hat{u}_{1,o}^{2i-2}\oplus \hat{u}_{1,e}^{2i-2}),L_{\frac{N}{2}}^{(i)}(y_{\frac{N}{2}+1}^{N}, \hat{u}_{1,e}^{2i-2}))
	\end{split}
\end{equation}
where $f(a,b)=ln(\frac{1+e^{a+b}}{e^{a}+e^{b}})$, and $g(a,b,c)=(-1)^{c}a+b$. $y_1^N$ is the received vector.

\subsection{SCL-Flip Decoders for Polar Codes}
To further improve the error-correction performance of CA-SCL decoder, some SCL-Flip decoders were successively proposed. These SCL-Flip decoding processes always contain two steps. 
The first step is an initial decoding attempt, which is almost consistent with the CA-SCL decoding attempt. But the initial decoding attempt will pass to the second step if the CRC in initial decoding attempt fails.

The second step contains a maximum of $T_{max}$ re-decoding attempts. Each re-decoding attempt has the same pruning rule on bit-channels as the first step, except a flip operation on a selected error-prone bit-channel recorded in a critical set (CS). It is noted that the flip operation is essentially one type of pruning rule. The second step will preferentially perform a re-decoding attempt that performs a flip operation on the bit-channel with a lower index in the CS. Then, the same CRC is checked at the end of each re-decoding attempts. The second step continues for $T_{max}$ attempts, or until the CRC passes.

Based on the above descriptions, the main difference between SCL-Flip decoders and CA-SCL decoder is extra re-decoding attempts that include flip operations. These flip operations on error-prone bit-channels are beneficial for improving the error-correction performance of CA-SCL decoder. Moreover, the performance improvement caused by flip operations depends on the CS and the detail of flip operations. CS records the location of flip operations, its size is $T_{max}$, and $CS \subseteq \mathcal{A}$.

The ways in generating CS among SCL-Flip decoders are different. SCL-Flip decoders in \cite{[15],[16],[17]} generate the CS offline based on the estimation of error-prone bit-channels in SC decoder, while SCL-Flip decoder in \cite{[18]} generates the CS online based on the effect of pruned candidate paths in SCL decoder.  

The flip operations among SCL-Flip decoders are also different. To show the difference, we showed different decoding attempts in Fig. 1. $PC(4, 3)$ with $L=4$ is adopted. Assume $\{u_1,u_3,u_4\}\subseteq\mathcal{A}$, $\{u_2\}\subseteq\mathcal{A}^c$, and the $4^{th}$ bit-channel is the error-prone bit-channel. Fig. 1(a) refers to the decoding attempt of the conventional CA-SCL decoder. Fig. 1(b) and Fig. 1(c) refer to re-decoding attempts in different SCL-Flip decoders, and these re-decoding attempts will perform a flip operation on the $4^{th}$ bit-channel. Black nodes represent the reserved paths and white nodes denote the pruned paths. The solid black line means the current $\hat{u}$ is 1, while the dotted black line means $\hat{u}$ is 0. For example, the blue line with an arrow in Fig. 1(a) implies a survived candidate path $\{\hat{u}_1, \hat{u}_2, \hat{u}_3, \hat{u}_4\}=\{1, 0, 0, 0\}$.

If CA-SCL only reserve a single descendant path from a node of $3^{rd}$ bit-channel like the survived path $\{\hat{u}_1, \hat{u}_2, \hat{u}_3, \hat{u}_4\}=\{0, 0, 1, 0\}$ in Fig. 1(a), the flip operation in Fig. 1(b), used in \cite{[15]}, will eliminate the single descendant path and then keep another descendant path from the same node of $3^{rd}$ bit-channel like $\{\hat{u}_1, \hat{u}_2, \hat{u}_3, \hat{u}_4\}=\{0, 0, 1, 1\}$. If not, the flip operation in Fig. 1(b) will reserve the same candidate paths as that in Fig. 1(a). 

Additionally, the flip operation in Fig. 1(c), used in \cite{[16],[17],[18]}, will keep all pruned paths on the $4^{th}$ bit-channel in Fig. 1(a).

It is noted that these SCL-Flip decoders related to this work all perform no more than one flip operation at each re-decoding attempt. The reason is that performing more than one flip operation at a re-decoding attempt slightly increases the error-correction performance of the above SCL-Flip decoders at the cost of a significantly increased computational complexity \cite{[18]}.
\begin{figure}[t]
	\centering
	\includegraphics[scale=0.6]{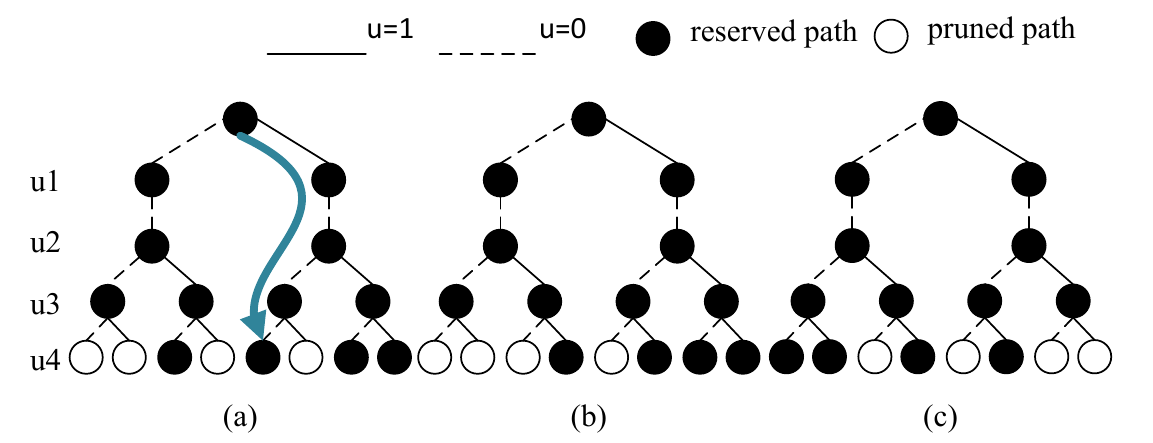}
	\caption{Decoding attempts for different list decoders, for $PC(4,3)$ with $L=4$.}
	\label{fig:1}       
\end{figure}

\section{AD-SCLF Decoder}\label{sec3}
Although SCL-Flip decoders improve the error-correction performance of CA-SCL decoder, the introduced flip operations result in redundant re-decoding attempts, especially at low to medium SNR. Moreover, these redundant re-decoding attempts increase decoding computation complexity and then require extra energy for decoding. To significantly reduce the computation complexity caused by redundant re-decoding attempts, we proposed an AD-SCLF decoder that can be implemented with an uncomplicated structure.

\subsection{Optimizing Indicators of Affecting Computation Complexity}

Based on the descriptions in Section \uppercase\expandafter{\romannumeral2}. C, the computation complexity of SCL-Flip decoders mainly depends on some indicators, like the efficiency of flip operation, the size of list, and the $T_{max}$ value. Thus, our AD-SCLF would reduce their computation complexity based on the efficiency of flip operation and the size of list. 

An efficient flip operation can be obtained by an efficient flip method and a good CS that records the location of flipping.

A more efficient flip method has a larger potential to improve the error-correction performance. For example, the flip method in Fig. 1(c) is more efficient in enhancing error-correction performance than that in Fig. 1(b), which was proved by the simulation results in \cite{[16]}.

Meanwhile, the CS in \cite{[18]} is more efficient in finding error-prone bit-channels than CS in \cite{[15],[16],[17]}. The reason is \cite{[18]} considers the effect of pruned candidate path in CA-SCL decoding process, while \cite{[15],[16],[17]} establish the CS based on the analysis to SC decoder. Reference \cite{[18]} designs a new threshold $E$ to generate the CS. $E$ is defined as
\begin{equation}
	E_{i}=log_{10}\frac{\sum_{l=1}^{L}e^{-PM_l^{(i)}}}{(\sum_{l=1}^{L}e^{-PM_{l+L}^{(i)}})^\alpha}
\end{equation}
where $E_i$ refers to the $E$ value of $i^{th}$ bit-channel. $\alpha$ is set as 1.2 according to \cite{[18]}. Obviously, as the $E_i$ decreases, the $i^{th}$ bit-channel is more likely to incur errors. Thus, the CS in \cite{[18]} is consists of indexes of bit-channels with smaller $E_i$ values, while the size is $T_{max}$, and $CS \subseteq \mathcal{A}$. 

The computation complexity is also severely affected by the list size. For example, the computation complexity of a re-decoding attempt in SCL-Flip decoders is $O(LNlogN)$, where $L$ is the current list size in the re-decoding attempt. Moreover, a list decoder has a big probability of obtaining the correct path with a small list size when SNR is high. Meanwhile, SCL-Flip decoders are essentially re-decoding algorithms. Therefore, an adaptive list is beneficial for reducing the computation complexity.  

\subsection{Details of AD-SCLF Decoder}

Based on the above analysis, we have designed a AD-SCLF decoder. Due to the flip efficiency, the flip method in Fig. 1(c) is adopted by our AD-SCLF decoder.  

Algorithm 1 shows the detail of AD-SCLF decoder. $y_1^N$ refers to the received vector. $L_{max}$ represents the maximum value of list size. $T_{max}$ means the maximum number of flip operations. As observed in Algorithm 1, the AD-SCLF contains two steps.

The first step contains multiple CA-SCL decoding attempts with different list sizes. $\hat{u}_1^N=\text{CA-SCL}(y_1^N, \mathcal{A}^C, K, L)$ means $\hat{u}_1^N$ is obtained by performing a CA-SCL decoder with paraments: $y_1^N$, $\mathcal{A}^C$, $K$ and $L$. $CRC(\hat{u}_1^N)=success$ represents the CRC of $\hat{u}_1^N$ successes. All CRCs in an AD-SCLF decoding are unchanged. The first step continues for $log_2(L_{max})$ attempts, or until the CRC passes. If all CRCs in the first step fail, the AD-SCLF will perform the second step. If not, the AD-SCLF will output the $\hat{u}_1^N$ that satisfies $CRC(\hat{u}_1^N)=success$ as the final result and then end the AD-SCLF decoding process.

The second step includes a CA-SCL decoding attempt and a maximum of $T_{max}$ SCL-Flip re-decoding attempts. $L_{max}$ is the list size in these attempts. $\hat{u}_1^N =\text{SCL-Flip}(y_1^N, \mathcal{A}^C, K, L_{max}, CS(t))$ means that $\hat{u}_1^N$ is obtained by performing an SCL-Flip re-decoding attempt with paraments: $y_1^N$, $\mathcal{A}^C$, $K$, $L_{max}$ and $CS(t)$.  $CS(t)$ refers to the $t^{th}$ elements in CS, which also is the location of flip at this re-decoding attempts. If all CRCs in the second step fail, the AD-SCLF will output the $\hat{u}_1^N$ in the final attempt as the final result and then end the AD-SCLF decoding process. If not, the AD-SCL will output the $\hat{u}_1^N$ that satisfies $CRC(\hat{u}_1^N)=success$ as the final result and then end the AD-SCLF decoding process.

It is emphasized that the purpose of our decoder is that significantly reducing the computational complexity of current SCL-Flip decoders at a practical FER.

\begin{algorithm}
\caption{AD-SCLF decoder}\label{algo1}
\begin{algorithmic}[1]
\Require$y_1^N$, $\mathcal{A}^c$, $K$, $L_{max}$, $T_{max}$
\Ensure $\hat{u}_{1}^{N}$ 
\State $l_{max}=log_2(L_{max})$
\State $crcPass\leftarrow false$
\For {$l\leftarrow$ 0 to $l_{max}$}
\State $L=2^l$
\If{$L\neq L_{max}$}
\State $\hat{u}_1^N=\text{CA-SCL}(y_1^N, \mathcal{A}^C, K, L)$.
\If{$CRC(\hat{u}_1^N)=success$}
\State $crcPass\leftarrow true$
\State				break.
\EndIf
\Else
\For{$t\leftarrow$ 0 to $T_{max}$}
\If{$t=0$}
\State $\hat{u}_1^N$=CA-SCL($y_1^N$, $\mathcal{A}^C$, $K$, $L_{max}$), while achieving CS. The size of CS is $T_{max}$.
\Else
\State $\hat{u}_1^N$ =SCL-Flip($y_1^N$, $\mathcal{A}^C$, $K$, $L_{max}$, $CS(t)$).
\EndIf
\If{$CRC(\hat{u}_1^N)=success$}
\State $crcPass\leftarrow true$
\State break.
\EndIf
\EndFor
\EndIf
\EndFor
\State \Return{$\hat{u}_1^N$}
\end{algorithmic}
\end{algorithm}

\section{Simulation Results}\label{sec4}
In this section, we compare the computational complexity and error-correction performance of our AD-SCLF algorithm with that of current SCL-Flip algorithms. This paper adopts binary additive white Gaussian noise channel (BAWGNC) and binary phase-shift keying (BPSK) modulation. All compared algorithms have the same generator polynomial of CRC  $g(x)=x^{16}+x^{15}+x^2+1$. The Gaussian approximation (GA) construction algorithm \cite{[21]} was adopted with a design-SNR of $4dB$. It is noted that the $T_{max}$ value and the error-prone indexes set of SCL-Flip decoder in \cite{[17]} are preset by offline simulation, and these parameters will change with different code rates. Therefore, all compared decoders adopt the same $T_{max}$ as that of the SCL-Flip in \cite{[17]}, for fair reasons.

\subsection{Error-correction Performance}

Fig. 2 plots the FER curves for different SCL-Flip algorithms: it can be seen the FER curves of all SCL-Flip algorithms almost overlaps at the same maximum list size. These results imply that our decoder has a similar error-correction ability with current SCL-Flip decoders at the same list size.
\begin{figure}[t]
	\centering
	\includegraphics[trim=35 10 20 30,clip,scale=0.4]{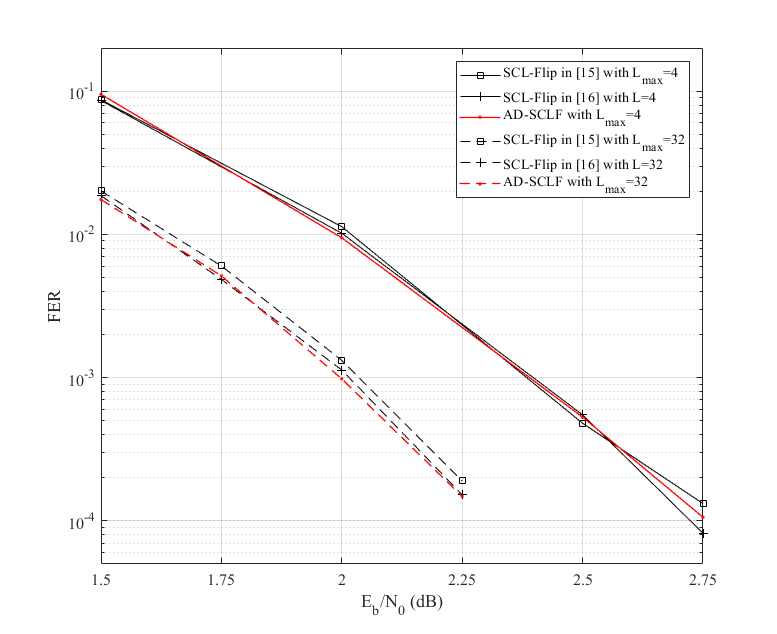}
	\caption{FER curves for $PC(512, 256+16)$ with $T_{max}=78$.}
	\label{fig:2}       
\end{figure}

\subsection{Computational Complexity}
Based on the previous analysis, an SCL-Flip decoding process is consists of an initial decoding attempt and some re-decoding attempts. To easily understand the computational complexity of SCL-Flip decoder, the initial decoding attempt and these re-decoding attempts are called decoding attempts in the following. Obviously, the computation complexity of an SCL-Flip decoder is the sum of the computation complexity of all attempts in this SCL-Flip decoding process. Meanwhile, the computation complexity of a decoding attempt is $O(LNlogN)$, where $L$ is the current list size in this decoding attempt. Thus, the computation complexity of SCL-Flip decoders can be represented as
\begin{equation}
	C_{sum}=\sum_{i=1}^{s}O(L_iNlogN)
\end{equation}
where $s$ is the sum of decoding attempts and $L_i$ represents the list size in the $i^{th}$ decoding attempt. Normalized with the computational complexity of SC decoder $O(NlogN)$, the representation of $C_{sum}$ can be simplified as
\begin{equation}
	c_{s}=\sum_{i=1}^{s}L_i
\end{equation}
where $c_s$ is directly proportional to $C_{sum}$.  In this section, $c_s$ is used for computation complexity comparison.

\begin{figure}[t]
	\centering
	\includegraphics[trim=35 10 0 30,clip,scale=0.4]{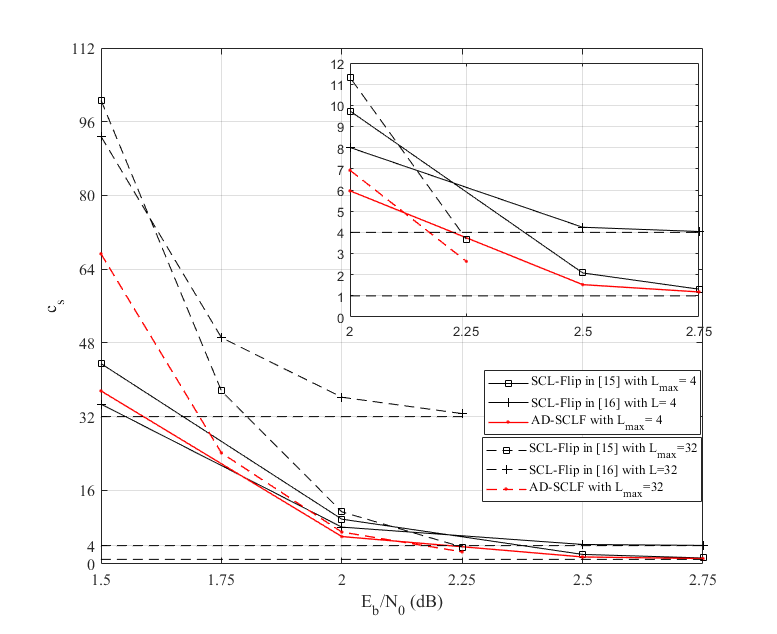}
	\caption{$c_s$ for $PC(512, 256+16)$ with $T_{max}=78$.}
	\label{fig:3}       
\end{figure}
Fig. 3 plots the $c_s$ for different SCL-Flip decoders: it can be seen that the AD-SCLF ($L_{max}=4$) has up to 70.69\% lower $c_s$ than SCL-Flip ($L=4$) in \cite{[18]} when $E_b/N_0 = 2.75dB$, and AD-SCLF ($L_{max}=32$) has up to 91.94\% lower $c_s$ than SCL-Flip ($L =32$) in \cite{[18]} when $E_b/N_0 = 2.75dB$. Meanwhile, AD-SCLF ($L_{max}=32$) reduces 80.85\% of the computation complexity of the SCL-Flip ($L =32$) decoder in \cite{[18]} at a matched practical $FER=10^{-3}$. The reason is that a candidate path passing the CRC can be obtained by an AD-SCLF decoder with a smaller list size than SCL-Flip decoder in \cite{[18]} at high SNR.  

Meanwhile, the AD-SCLF ($L_{max}=4$) has up to 38.76\% lower $c_s$ than SCL-Flip ($L_{max}=4$) in \cite{[17]} when $E_b/N_0 = 2dB$, and AD-SCLF ($L_{max}=32$) has up to 38.74\% lower $c_s$ than SCL-Flip ($L_{max}=32$) in \cite{[17]} when $E_b/N_0 = 2dB$. The reason is that the method of flipping in AD-SCLF decoder computes the effect of pruned candidate path in CA-SCL decoder, and then the AD-SCLF decoder is more efficient in finding the correct candidate path than that in SCL-Flip decoder in \cite{[17]}.

However, the computation complexity of AD-SCLF ($L_{max}=4$) has 8.2\% higher $c_s$ than SCL-Flip ($L_{max}=4$) in \cite{[18]} when $E_b/N_0 = 1.5dB$. The reason is that: on the one hand, a small list size, like $L_{max}=4$, increases the probability of eliminating the correct candidate path in the list decoding process for polar codes. On the other hand, the probability of multiple errors in a decoding attempt increase as $E_b/N_0$ decrease. Thus, the AD-SCLF decoder with a small list size may have a larger computational complexity than the SCL-Flip decoder in \cite{[18]} at low to medium $E_b/N_0$. But it is worth emphasizing that the computation complexity of AD-SCLF decoder is significantly lower than that of SCL-Flip decoders in \cite{[17],[18]}, at a practical $FER=10^{-3}$. Especially, the computational complexity of the AD-SCLF decoder is always lower than that of the SCL-Flip decoder in \cite{[17]}.

\section{Conclusion}\label{sec5}
To reduce the computation complexity of current SCL-Flip decoders, we have analyzed the indicators that affect the computation complexity. Based on these analyses, we have designed an AD-SCLF decoder, which can be easily implemented based on existing SCL-Flip structures. Simulation results demonstrate that the AD-SCLF decoder can significantly reduce the computation complexity of current SCL-Flip decoders at a matched practical FER.

\bmhead{Acknowledgments}

The authors acknowledge the support of the National Key R\&D Program of China under Grant (2021YFF0307602).

\section*{Declarations}

The authors have no relevant financial or non-financial interests to disclose.

\scriptsize
\bibliographystyle{unsrt}
\bibliography{reference.bib}
\balance



\end{document}